\documentclass[]{pasj02} 
\usepackage[switch,mathlines]{lineno} 

\usepackage{newtxtext,newtxmath}
\usepackage{ae,aecompl}
\usepackage{natbib}
\usepackage{graphicx}	
\usepackage{amsmath}	

\usepackage[dvipsnames]{xcolor}
\usepackage{multirow} 
\usepackage{comment}
\usepackage{xcolor}
\newcommand\msun{{\,M_\odot}}

\usepackage{adjustbox}

\newcommand{\Msun}{~\mbox{M}_{\odot}}

\newcommand{\LightGBM}{{\sc LightGBM}}
\newcommand{\ASloth}{{\sc A-SLOTH}}
\newcommand{\Cater}{{\sc Caterpillar}}

\newcommand{\nspace}{{ }}

\jyear{2025}
\Received{2025/09/10}
\Accepted{2025/11/28}

\begin{document} 

\title{Predicting Infall Time of Milky-Way Satellites via Machine Learning}

\author{
 Seungyeon \textsc{Kim}\altaffilmark{1}\orcid{0009-0008-7309-0840}
 ,
 Myoungwon \textsc{Jeon}\altaffilmark{1,2}\orcid{0000-0001-6529-9777} \email{myjeon@khu.ac.kr}
 , and
 Seongjun  \textsc{Hyung}\altaffilmark{1}\orcid{0009-0000-2881-9294}
}
\altaffiltext{1}{School of Space Research, Kyung Hee University, 1732 Deogyeong-daero, Giheung-gu, Yongin-si, Gyeonggi-do 17104, Republic of Korea}
\altaffiltext{2}{Department of Astronomy \& Space Science, Kyung Hee University, 1732 Deogyeong-daero, Yongin-si, Gyeonggi-do 17104, Republic of Korea}



\KeyWords{galaxies: dwarf, Local Group, galaxies: star formation, galaxies: evolution} 

\maketitle

\begin{abstract}
The properties of dwarf galaxies provide essential insight into galaxy formation and evolution in a hierarchical universe. Among various physical quantities, identifying their infall times to host galaxies is crucial, as these times encode key information such as star formation histories. However, estimating infall times remains challenging due to the complex interplay between different physical processes and the lack of consensus among existing methods. We propose a fast and interpretable method to predict the infall time of dwarf satellites using \LightGBM, a gradient-boosting decision tree algorithm. Our model is trained on satellites from 30 Milky Way (MW)-like host galaxies generated by \ASloth, a semi-analytic model calibrated using observational constraints, including those from the MW and its satellites. To balance predictive ability and observational applicability, we adopt $\tau_{90}$, [Fe/H], and $M_{\star}$ as input features. Since satellites with prior group membership hinder accurate MW infall predictions, we exclude them from the training data. As a result, the model achieves the best average mean squared error (MSE) of 5.04 in the \ASloth data set. Our model also shows good agreement with existing observational studies of MW satellites, although some discrepancies remain due to a few outliers such as CVn II and UMa I. In addition, for satellites experiencing prior infall events before MW-like host infall, the model predicts the timing of the first infall with a significantly lower MSE of 1.66, indicating the importance of the earliest infall in the quenching process of satellite galaxies.
\end{abstract}


\section{Introduction}


The star formation history (SFH), a key property of dwarf satellite galaxies, is shaped by various physical processes.
These processes include stellar feedback, especially supernova (SN) feedback, photoionization heating from stars, cosmic reionization, and various environmental effects (see \citealp{Tolstoy2009, Simon19} for reviews). When a dwarf galaxy crosses the virial radius of a more massive halo, it undergoes significant environmental influences, such as ram pressure and tidal stripping, which can substantially suppress star formation activity (e.g., \citealp{  Mayer2006, Wetzel13, Fattahi2018, Engler23}), often resulting in truncated SFHs. Thus, identifying the infall time of satellite dwarf galaxies into larger systems is crucial for understanding the factors that govern their SFHs.

\par
Despite their significance, there are limited methods for determining the infall times of satellite galaxies. Approaches involve calculating complete orbital histories via backward integration to extract infall times (e.g., \citealp{Petal2020, Miyoshi20, Benjamin21, Battaglia22, Bennet24, Santistevan2024, Taibi24}) and matching infall times from observational data with cosmological simulations (e.g., \citealp{Rocha12, Fillingham19}). Among these, backward integration from the initial proper motions of satellite galaxies is the most widely used technique for obtaining orbital histories and infall times. However, this method has certain limitations. Calculating orbits through backward integration is computationally intensive, leading many works to use simple spherically symmetric galaxy potentials, such as the Navarro-Frenk-White (NFW) profile (\citealp{Navarro96}). However, \citet{D’Souza2022} pointed out that these simple parametric potentials lack the sophistication needed to accurately reproduce the complex physical properties of host galaxies. In particular, low-mass satellites with shallow potential wells (typically \(M_{\star} \leq 10^6\msun\)) are particularly vulnerable to the host structure (e.g., \citealp{Akins21, Mercado25}).

\par


To address this limitation, various studies have focused on improving the simple potential of host halos. For example, \citet{Petal2020} emphasized the influence of the most massive satellites, such as the Large Magellanic Cloud (LMC), in describing the MW potential. They employed a combined potential of the MW and Magellanic Clouds (MCs) to demonstrate its impact on the orbital histories of satellite galaxies. On the other hand, \citet{Santistevan2024} tackled the limitations of using a fixed symmetric spherical host mass profile by adopting a static axisymmetric host potential to calculate the orbital histories of MW-like satellites. Despite these efforts, inferring infall times remains uncertain because orbits have often been integrated over the past $\sim$6 Gyr (e.g., \citealp{Patel2017, Petal2020, D’Souza2022}), without accounting for MW mass evolution and tidal mass loss. Since MW-mass galaxies have acquired about 80\% of their mass by 6 Gyr ago (e.g., \citealp{Santistevan2020}), integrating beyond this period requires a more complex orbital model that considers MW mass evolution and satellite mass loss.

To overcome the computational limitations in inferring infall times of dwarf galaxies, Machine Learning (ML) could offer an alternative solution. ML is a powerful tool for uncovering unknown relationships between different physical parameters. In modern astronomy, ML has addressed complex problems in many studies (see \citealp{Baron2019} for review), such as galaxy classification (e.g., \citealp{Zhu2019, Zeraatgari2024}), generating predicted observational data of the Sun (e.g., \citealp{Kim2019}), and predicting the mass of supermassive black holes (e.g., \citealp{agnet}). Furthermore, ML can discover entirely new relationships between previously unknown physical quantities (e.g., \citealp{Anders23}).

Regarding inferring infall times using ML techniques, \citet{Barmentloo23} predicted the infall times of the MW satellites using a Multi-Layer Perceptron (MLP). They utilized orbital phase-space information of satellite galaxies, including distance, radial velocity, and specific angular momentum, as input features. This MLP model allowed them to derive infall times directly from the proper motions of observational data. However, this approach is fundamentally similar to backward integration, as it also relies on the proper motions of the satellites at $z=0$. Consequently, \citet{Barmentloo23} faced the same limitation as backward integration: the inability to accurately predict infall times earlier than 6 Gyr ago. Moreover, \citet{Barmentloo23} did not explicitly account for whether a satellite had previously fallen into another host halo before its final infall onto the MW.


\par
Alternatively, one can estimate infall times by learning the correlations between infall times and related physical parameters. For instance, numerous studies have investigated the empirical relationship between infall times and the quenching times of satellite galaxies (e.g., \citealp{Wetzel13, Akins21,  Samuel22}). Among them, \citet{D'souza21} showed that classical spherical dwarfs with stellar mass of \(M_{\star} \sim 10^5 M_{\odot} - 10^7 M_{\odot} \) typically quench about 1-2 Gyr after infall into their hosts. Similarly, \citet{Fillingham15} demonstrated that low-mass satellites around MW-like host galaxies should be quenched within about 2 Gyr after infall to account for the high-quenched fraction observed in satellite galaxies. Note that quenching times are often derived from the cumulative star formation histories of observed galaxies, obtained by fitting the observed color-magnitude diagram (e.g., \citealp{Weisz14}).

In this study, we propose a new approach to efficiently estimate the infall times of satellites in MW-like galaxies by leveraging the correlation between infall times and quenching times. Specifically, if quenching times are available from observations, they can be used to infer infall times using \LightGBM, a tree-based ML algorithm (\citealp{Lightgbm}). Also, the relationship between quenching time and infall time may vary depending on whether dwarf galaxies underwent group preprocessing—falling into another halo—before infalling into the MW-like host galaxy. Therefore, we consider three scenarios to estimate infall times: (1) without considering group preprocessing, (2) considering only MW infall, and (3) considering the first infall into any host except the MW. We then compare the results across these conditions.



\par



\par
This paper is organized as follows. Section 2 introduces the training and test datasets generated by \ASloth~(\citealp{asloth, asloth2, Magg2022}), a semi-analytic model (SAM), along with the observational data used. We then describe the ML framework and model construction. In Section 3, we present and carefully analyze the results, including matching the model with observational data. Section 4 discusses the different model conditions and additional considerations, examining potential caveats. Finally, in Section 5, we summarize and conclude our work.

\begin{table}
\tbl{Summary of the data sets from \ASloth. Column (1): Name of the satellite group. Column (2): Stellar mass range of each group at $z = 0$. Column (3): Total number of satellites belonging to each group.}{%
\resizebox{0.9\linewidth}{!}{
\begin{tabular}{c|c|c}
\hline
Group & Mass Range [$\msun$] & Number of satellites \cr
\hline
{\sc Low} & $M_{\star} < 10^5$ & 11,708 \cr
{\sc Intermediate} & $10^5<M_{\star}\leq10^7$ & 308 \cr
{\sc Heavy} & $M_{\star} \ge 10^7$ & 154 \cr
\hline
\end{tabular}}}
\label{tab:data}
\end{table}

\section{Methodology} 
We aim to predict the infall times of satellite galaxies using machine learning techniques. To accomplish this, we construct a dataset and develop an appropriate predictive model. Section \ref{ss:data} details the data information and conditions utilized in this study. Section \ref{ss:model} provides a detailed description of the model construction process.

\subsection{Data}
\label{ss:data}
\subsubsection{Raw Data: \ASloth}
\label{subsubsec:asloth}
\par 

We extract data from \ASloth, a SAM that applies baryon physics to generate MW-like galaxies and their satellite dwarf galaxies, with a unique focus on the effect of metal-free and metal-poor stars in the early Universe (\citealp{asloth,asloth2}). \ASloth\ is calibrated using various observational constraints, including six MW-specific observables, and two cosmological relationships. This robust calibration ensures that the model provides a reliable representation of the real Universe. This framework enhances physical fidelity by incorporating sophisticated physics—including the effects of reionization on star formation in low-mass halos (discussed further in Section \ref{ss:ion})—while maintaining the computational efficiency of SAMs. Specifically, \citet{asloth} emphasized that \ASloth\ considers stochastic feedback from individual stars by tracking them individually. Thus, \ASloth\ is an ideal tool for providing samples of low-mass satellites ($M_{\star}\leq 10^4 \Msun$) generated with sophisticated subgrid physics.


\begin{figure*}
\begin{center}
\includegraphics[width = 180mm]{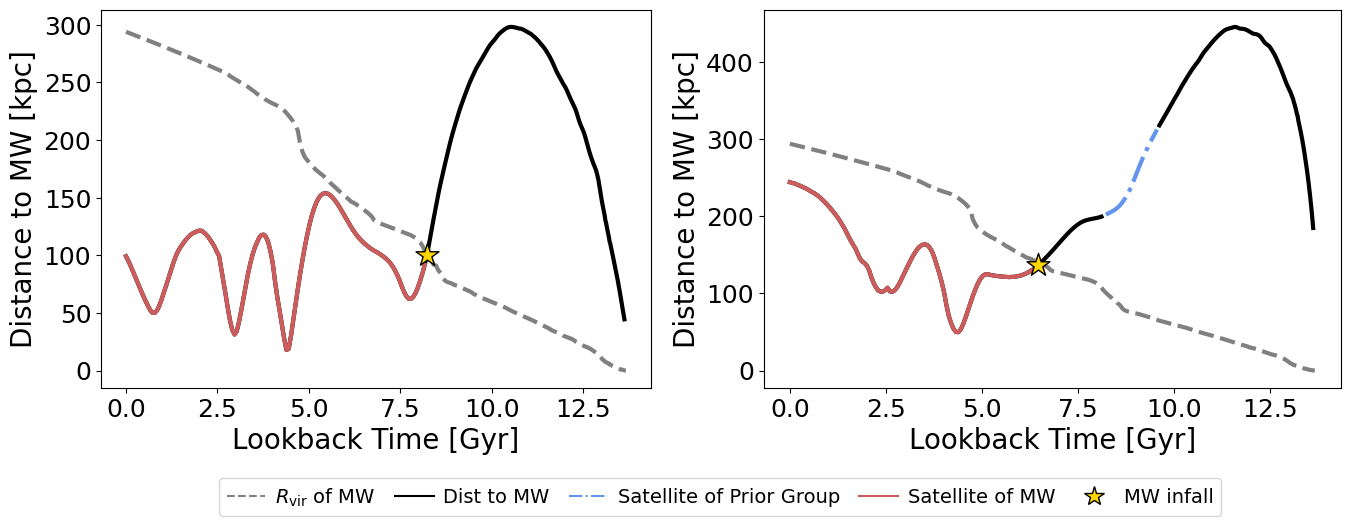}
\end{center}
    \caption{
    An example of the orbital history of a satellite galaxy is shown, depicting only infall into the MW-like galaxy (left) and first infall into a prior group (right) before infall into the MW-like galaxy in our \ASloth~data. The solid lines in both panels indicate the distance from the MW-like host galaxy to the satellites, while the gray dashed line shows the virial radius of the MW-like host galaxy. The blue dash-dot line in the right panel represents the period when the satellite is within the prior group, and the red lines depict periods when the satellite is inside the virial radius of the MW-like host galaxy. The infall time of the satellites, defined as the first passage of the satellite's center to the virial radius of the MW-like host galaxy, is marked by a yellow star.
    \\
    Alt text: Example of a satellite’s orbital history, illustrating a direct infall event onto the MW compared to a prior group infall event before reaching the MW.}
    \label{fig:fig1}
\end{figure*} 

\par 
To establish the model in \ASloth\, the flat $\Lambda$CDM Universe is assumed and cosmological parameters, obtained from Planck Collaboration, are used (\citealp{Plank}; $H_0 = 68 \mathrm{kms^{-1}~Mpc^{-1}}$, $\Omega_\mathrm{m} = 0.31$, $\Omega_\mathrm{\Lambda} = 0.69$, $\Omega_\mathrm{b}= 0.049$). \ASloth\ utilized 30 dark matter (DM) merger trees from the \Cater \ project (\citealp{Griffen16, Griffen18}). Specifically, each box size of the simulations extracted from the \Cater \ project has an effective side length ranging from 3 to 6 $h^{-1}$Mpc. For a more detailed description of the physical processes, we refer to \citet{asloth,asloth2}. Using 30 MW-like host halo merger trees, we generate a total of 12,170 satellite galaxies for this study. Since satellite galaxies exhibit varying physical properties depending on their stellar mass, we categorize them into three groups based on their stellar mass at $z=0$. Table~\ref{tab:data} presents the mass ranges and the number of satellites in each group. Most of the satellites belong to the low-mass group (96.2\%), followed by the intermediate-mass group (2.5\%) and the heavy-mass group (1.3\%).
This overabundance of satellite galaxies at the low-mass end naturally results from hierarchical structure formation. As shown in Figure.11a of \citet{Griffen16}, the subhalo mass function follows $dN/dM \propto M^{-1.88 \pm 0.10}$, indicating a steep increase in the number of subhalos at lower masses. Consistently, \citet{Griffen18} found that, although each of the 30 MW–like halos hosts on average 1,793 subhalos, only about 11 of them at $z=10$ have a maximum circular velocity with $V_\mathrm{max} > 50$ km s$^{-1}$.

\subsubsection{Data Preprocessing}
\label{sss: features and target}
\par

To predict the infall times of satellite galaxies, we select three features: (1) $\tau_{90}\ \mathrm{[Gyr]}$ -- the lookback time at which a galaxy formed 90\% of its total stellar mass, considered as the quenching time in this work, (2) $M_{\star}\ [\msun]$ -- the stellar mass of a satellite measured at $z = 0$, and (3) $\mathrm{[Fe/H]}$ -- the stellar metallicity of a satellite measured at $z = 0$. These features are chosen for their accessibility in observational data, facilitating application to observed MW satellite galaxies. Although additional properties such as magnitude or radial velocity could be used, they often correlate strongly with mass, potentially causing multicollinearity in the analysis.

\par

The infall time of a satellite is defined as the first instance when the distance between a dwarf halo and a host halo is less than the host halo's virial radius ($d_\mathrm{satellite,\ host}<R_\mathrm{vir,\ host}$). For detailed analysis, we classify infall times into two categories: the first infall ($t_{\rm first}$), which is the time when a dwarf galaxy first falls into any host halo excluding the MW, and the MW infall ($t_{\rm MW}$), which is the time when a dwarf becomes a MW satellite, either by directly infalling into the MW or after being part of a previous host halo.

\begin{figure} 
\begin{center}
    \includegraphics[width = 85mm]{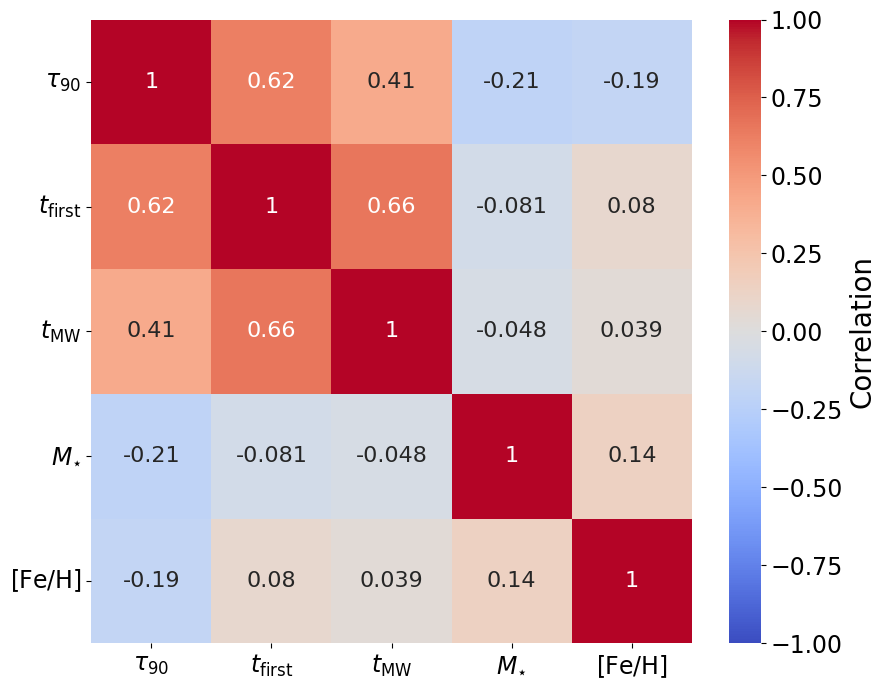} 
\end{center}
    \caption{Results of the Pearson correlation coefficient for each feature indicate that the first infall time, $t_\mathrm{first}$, has the strongest correlation with quenching time. The infall time to the MW-like host shows a moderate correlation, while metallicity ($\mathrm{[Fe/H]}$) and stellar mass ($M_{\star}$) of the satellites exhibit low correlation coefficients. This suggests that using $\mathrm{[Fe/H]}$ and $M_{\star}$ alone would result in inefficient model outcomes.
    \\
    Alt text: Correlation analysis of each feature with the first infall time onto any halo and the infall time onto an MW like galaxy.}
    \label{fig:corr}
\end{figure}

To determine these two types of infall times, we generate the orbital histories of all satellite galaxies. Figure \ref{fig:fig1} illustrates examples of two distinct satellite orbital histories. The solid line represents the distance to their MW-like host, while the dashed line shows the virial radius of an MW-like host galaxy, which increases as time progresses toward the present. The red and blue lines indicate periods when a galaxy belongs to the MW and a prior group, respectively. The left panel depicts a dwarf galaxy that directly fell into its MW-like host, whereas the right panel shows a dwarf galaxy that was part of a group before its infall into the MW-like host. We exclude dwarf galaxies that fell or quenched less than 0.1 Gyr ago. Dwarfs that entered and later escaped the MW, known as "backsplash galaxies," are also not included. Multiple infalls in the MW or other host halos are also ignored in our analysis. During this process, we removed 68 dwarf galaxies for which $\tau_{90}$ or infall times could not be computed. As a result, the final dataset used for analysis comprises 12,102 satellite galaxies.

\par 

We also compute the Pearson Correlation Coefficient (PCC; \citealp{Pearson96}) between infall times and each feature. A PCC of 1 indicates a perfect positive correlation, while a value of –1 indicates a perfect negative correlation. Figure \ref{fig:corr} shows the PCC values for each feature with respect to the two infall times ($t_{\rm MW},\ t_{\rm first}$). Among all features, quenching time exhibits the strongest correlation with infall time, whereas stellar mass and metallicity show relatively weak correlations.
Despite their weak individual correlations with the targets ($t_{\rm MW},\ t_{\rm first}$), these features ($M_{\star}$, $\mathrm{[Fe/H]}$) may still contribute to predictive performance through interactions with other variables. Furthermore, the tree-based models employed in this study are suited for capturing such complex nonlinear interactions. More details on this aspect are provided in Section \ref{sss:model}.


Note that while pericenter passages are known to be an efficient mechanism for quenching star formation in satellite galaxies (e.g., \citealp{Oman13, Wetzel15, Fillingham15}), the current A-SLOTH framework makes it challenging to track the detailed orbital evolution of satellites post-infall. Consequently, we cannot reliably estimate quantities such as the timing or strength of pericenter passages, as environmental processes like tidal and ram-pressure stripping are not explicitly modeled after infall (\citealp{asloth}). Consequently, we are unable to directly examine the correlation between $\tau_{90}$ and pericenter passage in this study. Future simulations that track full orbital dynamics will make such analyses feasible.

In addition, we test halo properties, such as halo mass and concentration, as alternative features to stellar mass. The correlation coefficients of these two quantities with infall times are about three to five times stronger than those of stellar mass. This likely reflects the fact that halo properties, being determined mainly by gravitational dynamics, are less affected by baryonic feedback processes such as reionization or supernovae. However, because halo mass and concentration are not directly observable in real systems, we do not include them in our final empirical model.
We also examine whether other SFH indicators, such as $\tau_{70}$ or $\tau_{50}$, show stronger correlations with infall events. However, we find that $\tau_{90}$ consistently exhibits the highest correlation with both infall times. This is likely because most satellites quench very rapidly, making $\tau_{70}$ and $\tau_{50}$ less meaningful measures of their star formation history.

\subsubsection{Observational Data}
\label{ss:obs}
\begin{table*}

\tbl{Summary of our selected features of the MW (left) and M31 (right) satellites. Column (1), (2) of each table: Name of the host and satellite galaxy. Column (3), (4), (5) of each table: $\tau_\mathrm{90}$ (\citealp{Brown14,Weisz15,Sacchi21,Bettinelli18}), $M_{\star}$ (\citealp{McConnachine12}), $\mathrm{[Fe/H]}$ (\citealp{Ho15, Battaglia22,Jennings23}, respectively.}{%
\renewcommand{\arraystretch}{1.2}
\makebox[\textwidth][c]{
\begin{tabular}{c|c|c|c|c||c|c|c|c|c}
\hline
\ \ \ Host\ \ \ \  & Satellite & $\tau_{90}$ [Gyr] & $M_{\star}$ [$M_\odot$] & [Fe/H] 
& \ \ \ Host\ \ \ \  & Satellite & $\tau_{90}$ [Gyr] & $M_{\star}$ [$M_\odot$] & [Fe/H] \\
\hline
\multirow{20}{*}{MW} & Boo I & 12.6 & $2.9\times10^4$ & -2.55 & \multirow{27}{*}{M31} & M32 & 3.1 & $3.2\times10^8$ & -1.11 \\
& Car I & 2.2 & $3.8\times10^5$ & -1.80 & & NGC 147 & 5.7 & $6.2\times10^7$ & -0.51 \\
& ComBer I & 13.0 & $3.7\times10^3$ & -2.00 & & NGC 185 & 9.3 & $6.8\times10^7$ & -0.98 \\
& CVn I & 8.3 & $2.3\times10^5$ & -1.98 & & NGC 205 & 2.6 & $3.3\times10^8$ & -0.87 \\
& CVn II & 12.7 & $7.9\times10^3$ & -2.21 & & And I & 9.0 & $3.9\times10^6$ & -1.56 \\
& Dra I & 9.1 & $2.9\times10^5$ & -2.10 & & And II & 7.6 & $7.6\times10^6$ & -1.52 \\
& Frn I & 2.4 & $2\times10^7$ & -2.26 & & And III & 10.1 & $8.3\times10^5$ & -1.78 \\
& Her I & 11.8 & $3.7\times10^4$ & -1.90 & & And V & 9.5 & $3.9\times10^5$ & -1.80 \\
& Hyd II & 2.2 & $7.1\times10^3$ & -2.02 & & And VI & 6.4 & $2.8\times10^6$ & -1.53 \\
& Leo I & 1.7 & $5.5\times10^6$ & -1.32 & & And VII & 7.7 & $9.5\times10^6$ & -1.34 \\
& Leo II & 6.4 & $7.4\times10^5$ & -1.60 & & And X & 7.5 & $9.6\times10^4$ & -2.03 \\
& Leo IV & 12.2 & $1.9\times10^4$ & -1.37 & & And XI & 12.6 & $4.9\times10^4$ & -2.14 \\
& Sag I & 3.4 & $2.1\times10^7$ & -2.10 & & And XII & 6.5 & $3.1\times10^4$ & -2.11 \\
& Scu I & 10.6 & $2.3\times10^6$ & -1.93 & & And XIII & 5.8 & $4.1\times10^4$ & -2.09 \\
& Sext I & 11.45 & $4.4\times10^5$ & -1.93 & & And XIV & 8.2 & $2.0\times10^5$ & -1.88 \\
& Tri II & 12.91 & $8.97\times10^2$ & -2.38 & & And XV & 9.4 & $4.9\times10^5$ & -1.91 \\
& Tuc II & 12.8 & $4.9\times10^2$ & -2.23 & & And XVI & 7.8 & $4.1\times10^5$ & -2.01 \\
& UMa I & 11.2 & $1.4\times10^4$ & -2.18 & & And XVII & 12.6 & $2.6\times10^5$ & -1.98 \\
& UMi I & 10.2 & $2.9\times10^5$ & -2.00 & & And XX & 8.3 & $2.9\times10^4$ & -2.14 \\
\cline{2-5}
& \multicolumn{4}{c||}{LMC Group} & & And XXI & 9.0 & $7.6\times10^5$ & -1.85 \\
\cline{2-5}
& Car II & 9.46 & $4.2\times10^4$ & -2.44 & & And XXII & 4.4 & $3.4\times10^4$ & -2.14 \\
& Hor I & 11.5 & $1.96\times10^3$ & -2.76 & & And XXIII & 6.6 & $1.1\times10^6$ & -1.74 \\
& Ret II & 12.29 & $1\times10^3$ & -2.46 & & And XXIV & 7.6 & $9.3\times10^4$ & -2.00 \\
& & & & & & And XXV & 10.2 & $6.8\times10^5$ & -1.82 \\
& & & & & & And XXVI & 6.3 & $6.0\times10^4$ & -2.18 \\
& & & & & & And XXVIII & 10.0 & $2.1\times10^5$ & -1.86 \\
& & & & & & And XXIX & 10.4 & $1.8\times10^5$ & -1.93 \\
\hline
\end{tabular}}}
\label{tab:obs_data}
\end{table*}

We apply our model to observed MW satellites to evaluate its performance under more realistic conditions. 
Although about 60 satellite galaxies have been discovered in the MW to date (e.g., \citealp{McConnachine12, Newton18, Simon19, Santos25, DD25}), only those with available measurements of $\tau_\mathrm{90}$, $M_{\star}$, and $\mathrm{[Fe/H]}$ are applicable in our model. Consequently, we select 22 MW satellites with all relevant data. In addition, we aim to predict the infall times of satellites associated with M31, another host galaxy in the Local Group, where the properties of satellite galaxies are observationally accessible. We compile the necessary observational data for 27 M31 satellites. Brief information on the data for the MW and M31 satellites is presented in Table \ref{tab:obs_data}.


\subsection{Model}
\label{ss:model}
\begin{figure}
\begin{center}
    \includegraphics[width = 85mm]{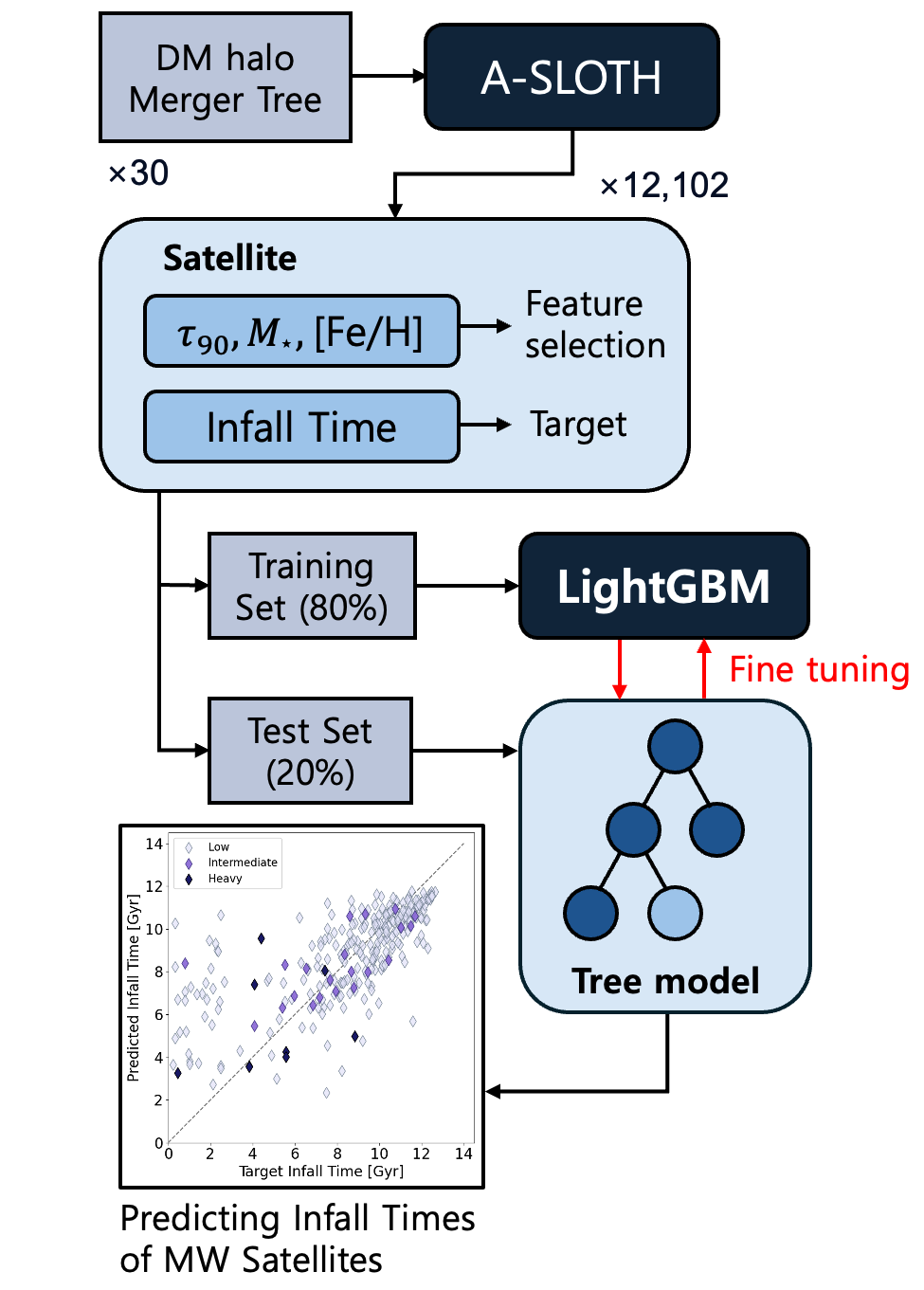}
\end{center}
    \caption{Flowchart for constructing the \LightGBM \nspace model. Utilizing 30 DM halo merger trees, we generate 12,102 satellite galaxies with \ASloth. From each satellite, we extract $\tau_{90}$ [Gyr], $M_{\star}\ [M_\odot]$, and $\mathrm{[Fe/H]}$ as features to predict the infall time.
    We randomly sample 80\% of the data for model training and optimize the model to find the best parameters, and the remaining 20\% is used to test the model to evaluate the performance.
    \\
    Alt text: Flowchart of the \LightGBM \ model, describing the process from data extraction for dwarf galaxies to training and testing.}
    \label{fig:flowchart}
\end{figure}


As detailed in Section \ref{subsubsec:asloth}, we utilize \ASloth\ to generate 12,102 satellite galaxies from 30 MW-like host halos. From the raw dataset, we compute infall times and select features such as $\tau_{90}$, $M_{\star}$, and $\rm [Fe/H]$ based on the criteria outlined in Section \ref{sss: features and target}. The dataset is randomly divided into training and testing subsets using \texttt{train\_test\_split} from \texttt{sklearn}, allocating 80\% for training and the remaining 20\% for testing. 
This module automatically shuffles the data so that the mass group ratios in the training and test sets remain representative of those in the entire dataset.
Figure \ref{fig:flowchart} presents the fundamental workflow for model construction. We introduce \LightGBM\ in Section \ref{LightGBM}, followed by a detailed explanation of the infall time prediction model and evaluation methodology in Section \ref{sss:model}.

\subsubsection{{\sc Light}GBM}
\label{LightGBM}
\par 

Gradient Boosting Machine (GBM) is an ensemble algorithm that sequentially combines multiple weak learners to create a robust model. It is trained to minimize the residual errors of the previous models through weight adjustments. GBM is known for its high predictive performance, particularly in handling nonlinear relationships and missing values. However, traditional GBM has several limitations, such as slow training due to its sequential nature and a propensity to overfit.

\par
\LightGBM, developed by Microsoft (\citealp{Lightgbm}), is an advanced gradient boosting framework that addresses these limitations. Unlike conventional decision tree algorithms, \LightGBM\ employs a leaf-wise growth strategy, which is effective for asymmetric data. It incorporates two key techniques: Gradient-based One-Side Sampling (GOSS), which enhances training efficiency by focusing on instances with large gradients, and Exclusive Feature Bundling (EFB), which reduces the dimensionality of sparse feature spaces by merging mutually exclusive features. These strategies enable \LightGBM\ to achieve faster training speeds and lower memory usage than traditional methods. In our study, we adopt the \LightGBM\ framework to predict satellite galaxy infall times, capitalizing on its computational efficiency and strong performance with complex, imbalanced data.

\subsubsection{Model Construction}
\label{sss:model}
\par
The hyperparameters of our model are empirically determined. Among the key parameters, the boosting type we use is Gradient Boosted Decision Tree (GBDT), which serves as the baseline model. We also experiment with other boosting types, but they do not provide any clear performance improvements over GBDT. The learning rate is set to 0.1, and the number of leaves is set to 31, which provided optimal performance during hyperparameter tuning. Model performance is evaluated using Mean Squared Error (MSE) as the loss function. To assess model stability and maximize data utilization, we apply 5-fold cross-validation. Given the strong mass asymmetry in our dataset, we also vary the number of training samples. While the model shows limited sensitivity to hyperparameter settings, it is notably dependent on the quantity of training data.

\par 
The selected features vary significantly in scale, with $10^{2} \leq M_{\star} [\msun]\leq10^{10}$ and $-3\leq\mathrm{[Fe/H]} \leq 0$, indicating varying orders of magnitude. As such, the large disparity in feature scales makes it difficult to fairly evaluate their importances and slows down the convergence. Therefore, feature scaling is required to ensure an unbiased evaluation. We scale our data with standardization using \texttt{StandardScaler} from \texttt{Scikit-learn}.
By using the \texttt{StandardScaler}, each feature is standardized to have zero mean and unit variance, thereby alleviating the large scale differences among features.

\section{Result}
\label{sec:3}
In this section, we present and analyze the results of our model for predicting the infall times of satellite galaxies. Section \ref{subsec:performance} explains the predictive performance of \LightGBM\ models in various combinations of features, discussing the most significant features that contribute to predictions. Sections \ref{subsec:obs} and \ref{subsec:M31} explore the application of our model to observed satellite galaxies in the MW and M31, respectively, comparing our results with previous studies.

\subsection{Model Performance}
\label{subsec:performance}

As shown in Section \ref{sss: features and target}, the correlation between the quenching time, $\tau_{90}$, and the first infall time ($t_{\rm first}$) is notably stronger than the MW infall time ($t_{\rm MW}$). Consequently, we construct two separate models: one using the full dataset and another with only MW first infall time information, to examine the impact on performance. In Section \ref{sss:firstinfall}, we discuss the preprocessing of groups and the role of first infall in satellite evolution. We then assess how model performance varies in Section \ref{sss:mse} and further analyze the importance of each feature in Section \ref{sss:fea_imp}.

\subsubsection{The group preprocessing}
\label{sss:firstinfall}
Group preprocessing refers to the process in which a galaxy becomes gravitationally bound to another more massive halo, joining its group, before falling into their host galaxy at z = 0 (e.g., \citealp{Fujita04, Wetzel13}).
During this process, some satellite galaxies leave their original group and fall into the current host galaxy individually, while others infall as part of the group. Empirically, several MW satellites—such as Car II, Hor I, Hyd I, and Phx II—are considered members of the LMC group (e.g., \citealp{Battaglia22}) before becoming satellite galaxies of the MW. This phenomenon is consistent with hierarchical cosmology, where smaller structures form first and later merge into larger systems.

\par 

Through group preprocessing, dwarf galaxies can experience various physical processes such as ram pressure stripping, tidal disruption, and the heating and ionizing effects of radiation from a host galaxy. These processes may quench star formation within a dwarf galaxy, complicating the identification of the correlation between quenching time and infall time to a MW-like host galaxy. Furthermore, the impact of group preprocessing is more pronounced in low-mass satellites, which are more vulnerable to environmental influences. Given that our dataset comprises a significant number of such low-mass galaxies, appropriate handling of this population is crucial for accurate modeling.

\begin{figure}
\begin{center}
    \includegraphics[width = 85mm]{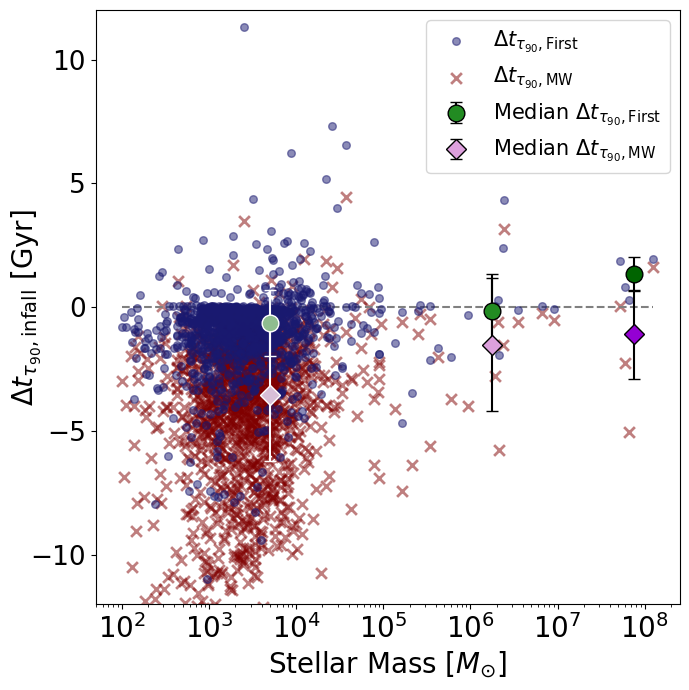}
\end{center}
    \caption{The quenching timescale of the first infall time (blue circle) and the MW infall time (red cross) of satellites which had belonged to prior groups. Green circles and purple diamonds represent the median timescale and 1-$\sigma$ uncertainty for the first infall and MW infall times, respectively. Color opacity indicates different mass groups, with lower opacity corresponding to lower-mass groups. Proximity to the gray dashed line indicates a smaller interval between the infall event and quenching time.
    \\
    Alt text: Quenching timescales of satellites related to infall events, derived from the \ASloth\ model.}
    \label{fig:f_i_M_i}
\end{figure}

\begin{table}
\tbl{The median, mean absolute value, and standard deviation of the quenching timescales of the first infall and the MW infall in each satellite mass group, in units of Gyr.}
{
\resizebox{0.95\linewidth}{!}{%
\begin{tabular}{c|c|c|c|c|c|c}
\hline
& \multicolumn{2}{c|}{Median}& \multicolumn{2}{c|}{Mean Absolute} & \multicolumn{2}{c}{$\sigma$} \cr
\hline
 & First & MW & First & MW & First & MW \cr
\hline
Low & -0.64 & -3.54 & 1.15 & 4.15 & 1.49 & 2.74 \cr
Intermediate& -0.15 & -1.52 & 1.34 & 2.68 & 1.91 & 2.61 \cr
Heavy & 1.34 & -1.12 & 1.22 & 2.25 & 0.69 & 2.51 \cr
\hline
\end{tabular}}
}
\label{tab:mean_std}
\end{table}

In our {\sc A-sloth} data, we find that 6,512 satellites (53.5\%) experience group preprocessing, while 5,590 satellites (46.5\%) infall directly into the MW-like host galaxy. Figure \ref{fig:f_i_M_i} represents the quenching timescale of the former satellites that experience group preprocessing before entering the MW-like galaxy. The quenching timescale, $\Delta t_{\tau_{90},\mathrm{infall}}$, is defined as the time difference between satellite infall and quenching, given by: $\Delta t_{\tau_{90},\ \mathrm{infall}} = t_\mathrm{infall}-\tau_{90}$. The blue circles and the red cross markers in Figure \ref{fig:f_i_M_i} indicate the quenching timescale of the first fall ($\Delta t_{\tau_{90},\ \mathrm{first}}$) and the MW infall ($\Delta t_{\tau_{90},\ \mathrm{MW}}$), respectively. We also show the median values and its 1-$\sigma$ uncertainty for each quenching timescale using green circles (first infall) and purple diamonds (MW infall). Marker opacity reflects different mass groups, with lower opacity indicating lower-mass satellites. In Table \ref{tab:mean_std}, we summarize the median, mean absolute, and 1-$\sigma$ value for each mass group.


As shown in Figure \ref{fig:f_i_M_i} and Table \ref{tab:mean_std}, the median of $\Delta t_{\tau_{90},\ \mathrm{first}}$ is shorter than $\Delta t_{\tau_{90},\ \mathrm{MW}}$ except for the case of heavy-mass group. Moreover, the mean absolute value of $\Delta t_{\tau_{90},\ \mathrm{first}}$ is also smaller, suggesting a closer connection between the first fall event and the quenching time, $\tau_{90}$. The difference between $\Delta t_{\tau_{90},\ \mathrm{first}}$ and $\Delta t_{\tau_{90},\ \mathrm{MW}}$ increases for lower-mass satellite groups, suggesting that the first infall event has a more pronounced effect on quenching in these satellites. Furthermore, the standard deviation of $\Delta t_{\tau_{90},\ \mathrm{MW}}$ is approximately twice that of $\Delta t_{\tau_{90},\ \mathrm{first}}$. Figure \ref{fig:f_i_M_i} supports this with a wide scatter in the MW infall data, implying that quenching is less sensitive to the MW infall event. Therefore, $\Delta t_{\tau_{90},\ \mathrm{MW}}$ appears to be less strongly associated with quenching compared to $\Delta t_{\tau_{90},\ \mathrm{first}}$ when satellites undergo group preprocessing before infalling into the MW-like galaxy.
This trend is also consistent with previous hydrodynamic simulation studies. For example, \citet{Samuel22} investigated the environmental quenching of satellite galaxies around MW-like hosts using the FIRE-2 simulations. They defined quenching timescales in a similar manner to examine the effect of group preprocessing and found that, for low-mass satellites with $M_* \lesssim 10^7\ M_\odot$, the quenching timescale after the first infall becomes shorter when group preprocessing occurs. This result aligns with our findings, further highlighting the importance of group preprocessing in shaping the quenching history of low-mass satellites.

\begin{figure*}[tp]
\begin{center}
    \includegraphics[width = 180mm]{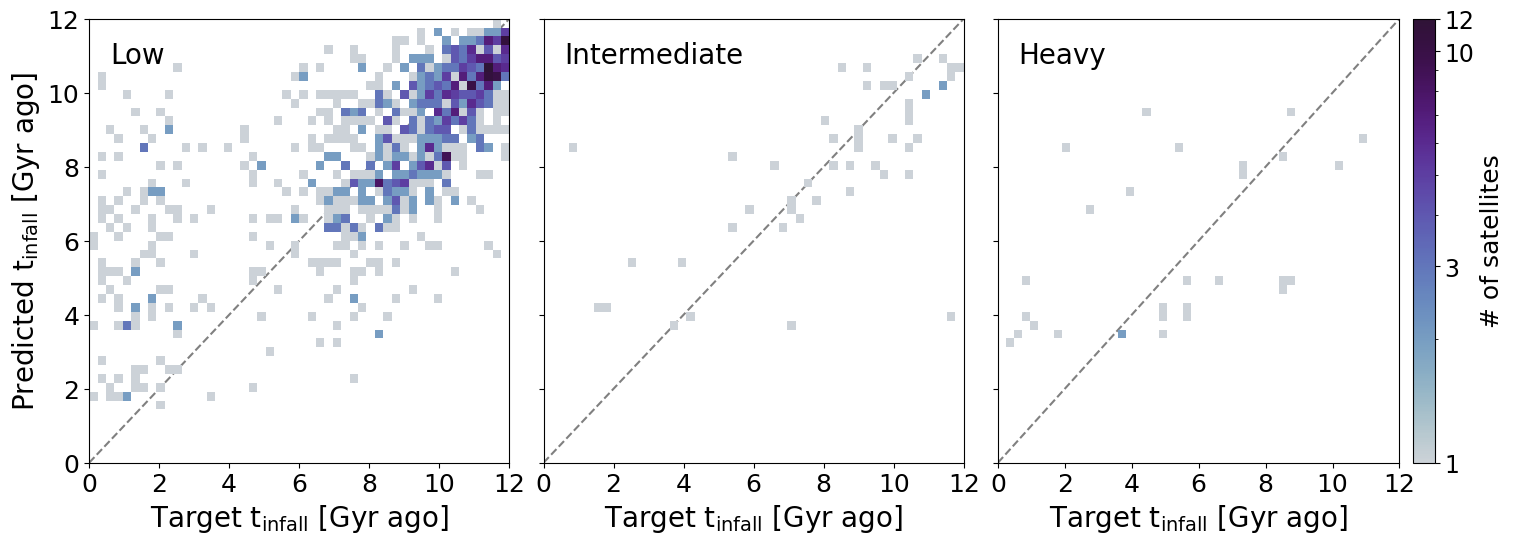} 
\end{center}
    \caption{Comparison between the target infall times of satellite galaxies from \ASloth~(considered as true values) and those predicted by our \LightGBM~model, shown in units of lookback Gyr as a heatmap. The left, middle, right panels correspond to low-, intermediate-, heavy-mass gorup, with the color intensity indicating the number of satellites at each point. Markers close to the dashed grey diagonal indicate a close match between predicted and target infall times. We find that the intermediate-mass group shows the best agreement, while the heavy-mass group exhibits the largest scatter. In the low-mass group, the scatter increases for satellites with infall times more recent than $\sim$ 4 Gyr ago.
    
    Alt text: Comparison of target and predicted infall times of \ASloth\ satellites across different mass groups.}
    \label{fig:pred-targ}
\end{figure*}

\subsubsection{MSE Loss of models}
\label{sss:mse}
\par
Under these circumstances, dwarfs that have belonged to prior groups may inhibit the ability of the model to predict the infall time of a MW-like galaxy. Therefore, we exclude these dwarfs and use only 5,590 dwarf galaxies without prior group association to train the model, although the first infall time is known to be strongly correlated with the quenching time. We find that the MSE loss decreases from 6.92 to 5.04 when using this selected dataset. In addition, we find that equalizing the sample size across groups reduces the loss to as low as 4.26. However, the total number of samples becomes too small ($\sim$ 500), resulting in fluctuations in the MSE loss depending on the validation set. As such, due to its instability, we do not adopt the model with equalized sample sizes among the three groups.

\setlength{\tabcolsep}{13pt} 
\begin{table}
\tbl{The importance of each feature. Row (1): each feature used to predict the infall times. Row (2): the information gain of each feature. Row (3): the split value of each feature.}{
\resizebox{0.85\linewidth}{!}{%
\begin{tabular}{c|c|c|c} 
\hline
Feature & $M_{\star}\ [M_\odot]$ & $\mathrm{[Fe/H]}$ & $\tau_\mathrm{90}\ \mathrm{[Gyr]}$ \cr
\hline
Gain & 16,322 & 20,058 & 93,640 \cr
Split & 394 & 401 & 405 \cr
\hline
\end{tabular}}}
\label{tab:fea_imp}
\end{table}

Figure \ref{fig:pred-targ} compares the target infall times of satellite galaxies with those predicted by our {\sc Light}GBM model. The mean MSE loss from 5-fold cross-validation is 5.04, with losses of 5.07, 4.15, and 5.28 for the low-, intermediate-, and heavy-mass groups, respectively. For the low-mass group, shown in the left panel of Figure \ref{fig:pred-targ}, the deviation is relatively large for satellites that infell within the last 4 Gyr, whereas earlier infalling satellites show better accuracy; the MSE for the relatively recently infalling 116 satellites is 23.29, which is about 9 times greater than that for the earlier infalling 922 satellites (2.54). This is because star formation in low-mass satellites is more susceptible to physical mechanisms such as cosmic reionization and stellar feedback, compared to host infall events in more massive satellites. In particular, during the early Universe, cosmic reionization is likely to uniformly truncate star formation in small galaxies (e.g., \citealp{Weisz14, Brown14}), and SN feedback in such galaxies easily quenched star formation by evacuating gas within their shallow potential wells (e.g., \citealp{Simpson2013, Jeon2017, Zhang2024}). Consequently, smaller galaxies tend to quench earlier in the Universe's history, a trend well-reflected in {\sc A-Sloth}. Therefore, if quenching occurred due to these mechanisms before experiencing host infall, the predictive power of the model for recently infalling low-mass galaxies declines, leading to a larger MSE.

\setlength{\tabcolsep}{3pt} 
\begin{table}
\tbl{The MSE loss for different feature combinations. The first row lists the model names, the second row displays various feature combinations, and the third row indicates the MSE loss for each condition.}{
\resizebox{1.01\linewidth}{!}{%
\begin{tabular}{c|c|c|c|c|c|c} 
\hline
Model & L1 & L2 & L3 & L4 & L5 & L6 \\
\hline
Feature & $M_{\star}$ & $\mathrm{[Fe/H]}$ & $\tau_\mathrm{90}$ & $M_{\star}$, $\mathrm{[Fe/H]}$ & $\mathrm{[Fe/H]}$, $\tau_\mathrm{90}$ & $M_{\star}$, $\tau_\mathrm{90}$\\
\hline
MSE loss & 9.24 & 9.47 & 6.06 & 9.03 & 5.48 & 5.53 \\
\hline
\end{tabular}
}}
\label{tab:fea_com}
\end{table}

Heavy-mass satellites, presented in the right panel of Figure \ref{fig:pred-targ}, are less affected by the ionizing or gravitational effects of the host galaxy, leading to a weaker correlation between infall and quenching times, and resulting in higher MSE values. This is because, in relatively massive galaxies, the central star formation regions are well-shielded, allowing stable star formation regardless of their entry into the host. On the other hand, the intermediate-mass group, shown in the middle panel of Figure \ref{fig:pred-targ}, shows the best agreement between predicted and target infall times.
\begin{figure*}
\begin{center}
    \includegraphics[width = 180mm]{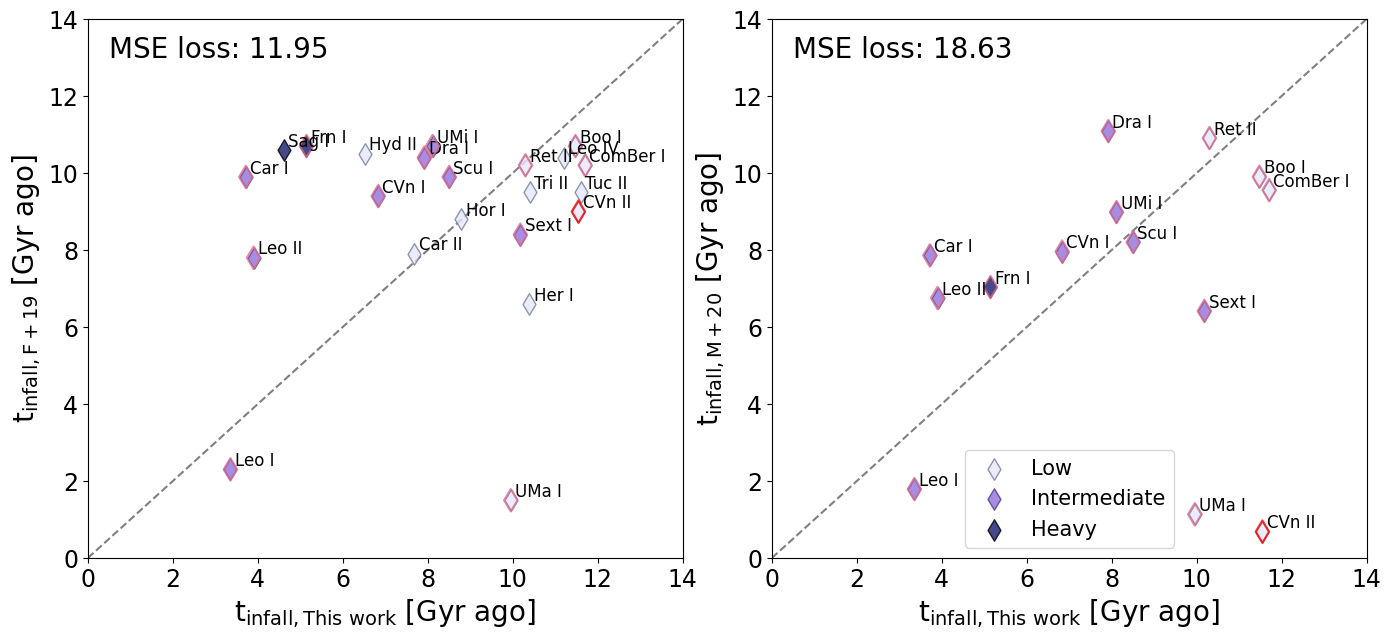}
\end{center}
    \caption{Comparison between the predicted infall times (in units of lookback Gyr) of MW satellite galaxies from this work and those reported in previous studies. The left panel compares our predictions with F+19, and the right panel compares with M+20. Each point represents a MW satellite galaxy and the color of each point is classified by the $M_{\star}$ of the galaxy at $z = 0$. Galaxies appearing in both F+19 and M+20 have pink edges, with CVn II highlighted by a red edge to indicate a significant discrepancy between studies. MSE losses are 11.95 and 18.63 for F+19 and M+20, respectively. While overall agreement is good, some scatter remains. The dashed gray line corresponds to the perfect prediction. This shows methodological differences among the studies and intrinsic uncertainties in infall time estimation.    
    \\
    Alt text: Comparison of predicted MW satellite infall times with those from previous studies.
    }
    \label{fig:obs_pred}
\end{figure*}

\subsubsection{Feature Importance}
\label{sss:fea_imp}
\par
We assess feature importance using gain and split, which represent the contribution and usage frequency of each feature, respectively. Split indicates the total number of times a feature is used as a splitting criterion in the decision tree, while gain measures the total information gain (IG) accumulated across all splitting points in the model. IG quantifies the reduction in entropy achieved by partitioning the data based on a given feature, making it a common criterion for determining splits in decision tree models. These measures are relative comparisons among features rather than absolute criteria. Generally, gain is considered the more influential metric.

\par
In Table \ref{tab:fea_imp}, we summarize the importance of each feature. Among the three input features, $\tau_{90}$ shows the highest total gain of 93,640, which is 4.67 times higher than the gain of $\mathrm{[Fe/H]}$ and 5.74 times greater than that of $M_{\star}$. Since gain is a relative measure of a feature’s contribution to reducing the model loss, we conclude that $\tau_{90}$ is the most influential feature for predicting the infall time of satellite galaxies. 
When compared with the Pearson correlation analysis presented in Section \ref{sss: features and target}, the results show consistent trends. For instance, $\tau_{90}$, which exhibits the strongest correlation with the infall time, displays a dominant gain value relative to the other features. In contrast, $\mathrm{[Fe/H]}$ and $M_{\star}$ yield similarly low gain values, as they show weak correlations with the infall time. This reinforces the strong physical connection between $\tau_{90}$ and the infall process. However, a high gain value does not imply that other features are unimportant, as they still provide complementary information that improves the overall model performance. The split counts for the three features are all similar in magnitude -- 394 for $M_{\star}$, 401 for $\mathrm{[Fe/H]}$, and 405 for $\tau_{90}$. This suggests that all features contribute meaningfully to model construction, as relatively few splits would indicate that they were largely ignored during tree building. The comparable split frequencies imply that each feature provides valuable complementary information in different regions of the parameter space.

Furthermore, we quantify the impact of each feature by evaluating the model performance under different combinations of features. The MSE loss values for each setting are presented in Table~\ref{tab:fea_com}. We first train models using each feature individually, labeled as L1, L2, and L3. Among them, the model using only $\tau_{90}$ (L3) yields the lowest MSE loss, reaffirming that $\tau_{90}$ is the most influential feature. However, using $\tau_{90}$ alone does not result in a sufficiently low loss compared to the value of the model using all features (5.04), suggesting that additional features are necessary to improve the performance of the model. A similar trend is shown when using feature pairs in L4, L5, and L6. The best performance is achieved with L5, which combines $\tau_{90}$ and $\mathrm{[Fe/H]}$. This indicates that while $M_{\star}$ and $\mathrm{[Fe/H]}$ may not be highly predictive on their own, they provide complementary information that improves the model when used alongside $\tau_{90}$.

\subsection{Comparison with Infall times of MW satellties} 
\label{subsec:obs}
There have been various attempts to estimate the infall time of MW satellites. For example, \citet{Fillingham19} (hereafter F+19) combined 6D phase-space information from Gaia DR2 with Phat ELVIS cosmological simulations to derive the infall times of MW satellite galaxies based on their current positions and velocities. Specifically, by comparing the binding energies of the observed satellites with the orbital properties of subhalos in the simulations, they identified the most similar subhalos and derived representative infall times for the observed satellites. Meanwhile, \citet{Miyoshi20} (hereafter M+20) utilized proper motion data from Gaia DR2 combined with an evolving galactic potential model to trace satellite galaxy orbits back 13.5 billion years. This approach reconstructed the early accretion history of dwarf galaxies by relaxing the static potential assumption, showing qualitative agreement with cosmological simulation results from F+19.

\setlength{\tabcolsep}{10pt} 
\begin{table}
\tbl{Comparison of MSE loss for MW satellite infall times between our model and B+23. The target values are based on infall times from F+19 and M+20. The MSE losses are calculated for different mass groups as well.}{
\resizebox{0.9\linewidth}{!}{%
\begin{tabular}{c|c|c|c}
\hline
\ \ Target\ \  & Mass group & This work &  \ \ \ B+23\ \ \   \\
\hline
\multirow{4}{*}{F+19} & Low & 9.75 & 20.33 \\
                      & Intermediate& 9.89 & 11.63 \\
                      & Heavy & 33.37 & 5.32\\
                      & Total & 11.95 & 15.80 \\
\hline
\multirow{4}{*}{M+20} & Low & 40.64 & 29.07 \\
                      & Intermediate& 6.23 & 6.52 \\
                      & Heavy & 3.57 & 1.35 \\
                      & Total & 18.63 & 14.20 \\
\hline
\end{tabular}}}
\label{tab:MFB}
\end{table}

\par 
Therefore, we compare the infall times predicted by our model with those reported in the two previous studies. Among the 22 MW satellite galaxies we compiled, all 22 overlap with the list from F+19, and 12 are included in the satellite list of M+20. Figure \ref{fig:obs_pred} shows the comparison with F+19 in the left panel and with M+20 in the right panel. The computed MSE losses are 11.95 for the comparison with F+19 and slightly higher at 18.63 for M+20. We should note that we directly use the suggested infall time values from F+19 and M+20. While our predictions generally agree well with both studies, some scatter is expected due to the different methodologies used in F+19 and M+20. Outliers such as UMa I and CVn II significantly increase the MSE loss. Excluding these satellites reduces the MSE for M+20 to 5.43, meaning that our model performs well for typical satellites aside from such outliers. For a more precise analysis, we classify the data by mass group and compute the MSE for each, as summarized in Table \ref{tab:MFB}. The MSE is relatively smaller in the low- and intermediate-mass groups compared to the heavy-mass group, suggesting that model performance varies with satellite mass. Compared to M+20, the large MSE of 40.64 in the low-mass group is mainly due to the aforementioned outliers. Removing them substantially reduces the MSE to 2.48.

\par 
We further compare our results with \citet{Barmentloo23} (hereafter B+23), where they predicted the infall times of MW satellite galaxies using MLP with 6-D phase space information as input features. Unlike F+19 and M+20, which do not use ML, B+23 is more directly comparable, given that both this work and B+23 estimate infall times using ML techniques. The comparison of MSE loss based on each mass group across the three previous studies is summarized in Table \ref{tab:MFB}. Since B+23 also provides predicted infall times for each satellite, we extract the infall time values of overlapping satellites from each mass group to compare with F+19 and M+20 and calculate the MSE. In the low-mass group, our model achieves a lower MSE compared to B+23 for F+19, but it results in a higher loss than B+23 for M+20. This discrepancy is likely due to CVn II, as its values differ significantly between F+19 and M+20. Our prediction matches well with F+19, but not with M+20, as indicated by the symbol with a red outline in Figure \ref{fig:obs_pred}, which contributes to a larger MSE compared to B+23.

For the intermediate-mass group, our model outperforms B+23 for both F+19 and M+20, while B+23 excels in the heavy-mass group. These results suggest that our model is more effective for low- and intermediate-mass satellites, whereas B+23 demonstrates better predictive ability for massive satellites. This may be attributed to the simulation resolution used for training the model in B+23 ($m_{\rm gas}\sim10^6\msun$), which could lead to better sampling of more massive systems than low-mass satellites. This is also evident in Figure 4 of B+23, where significant errors are found at epochs earlier than 6 Gyr ago -- when most satellites in the low- and intermediate-mass groups are accreted.


In general, while the performance of our model is similar to that of B+23, our approach offers a significant advantage in computational efficiency. B+23 used an MLP architecture with a relatively large number of features. In contrast, our model uses \LightGBM\ with fewer input features to achieve comparable performance. In particular, when tested under the same computational environment and features, we confirm that the MLP requires about 60 times more training time than \LightGBM\ for the same task. This efficiency makes our model particularly well-suited for large-scale applications that involve handling a large number of features.

\begin{figure}[tp]
\begin{center}
    \includegraphics[width = 85mm]{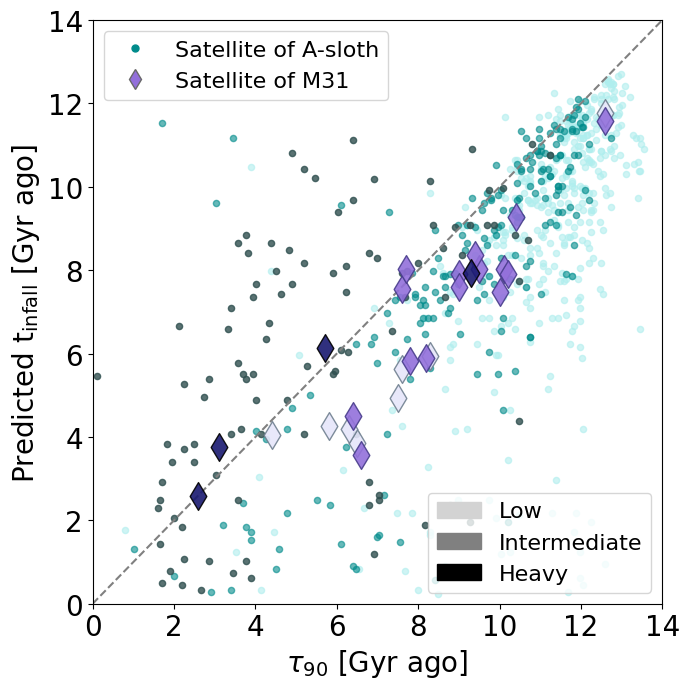}
\end{center}
    \caption{The predicted infall time of M31 satellite galaxies (in units of lookback Gyr) as a function of $\tau_{90}$. Each diamond marker shows the predicted M31 infall time, and the circles are randomly sampled satellites taken from \ASloth \nspace data. The predicted infall times well reproduce the trend of \ASloth \nspace satellite data in that they present high correlation with the $\tau_{90}$.
    \\
    Alt text: The infall times of M31 satellites predicted by our machine learning model.}
    \label{fig:M31}
\end{figure}

\subsection{Infall times of M31 satellites}
\label{subsec:M31}

We also predict the infall times of M31 satellites, as listed in Table 2, another Local Group galaxy comparable to the MW. In Figure \ref{fig:M31}, we plot the predicted infall times, represented by diamond markers, as a function of $\tau_{90}$. It is important to note that these are purely model predictions, as no previously calculated infall times for M31 satellites are available. Therefore, we overlay satellite galaxies extracted from {\sc A-Sloth} (shown as small circles in Figure \ref{fig:M31}) to examine whether our predictions exhibit a consistent trend. We randomly sample 300, 200, and 100 satellites from the low-, intermediate-, and heavy-mass groups, respectively. We do not compare our predictions with those of the MW satellites, as F+19 and M+20 reported different infall times for the same satellites, and neither can be rigorously regarded as the true value.

The predicted infall times for the M31 satellites successfully reproduce the existing trends from the data \ASloth \nspace, exhibiting a significant correlation with $\tau_{90}$. In particular, satellites belonging to the heavy-mass group tend to fall into the M31 before quenching, whereas those of the low- and intermediate-mass groups occur closely with $\tau_{90}$. This trend is also clearly shown in the \ASloth \nspace data. However, since these predictions are not directly compared with target infall times such as those from F+19 or M+20, variability may exist. As seen in the MW-like case, outliers, such as CVn II or UMa I, can cause significant deviations from trends. Therefore, once the infall times for the M31 satellites are available, we expect to better validate the predictability of our model.

\begin{figure}[tp]
 \begin{center}
  \includegraphics[width = 85mm]{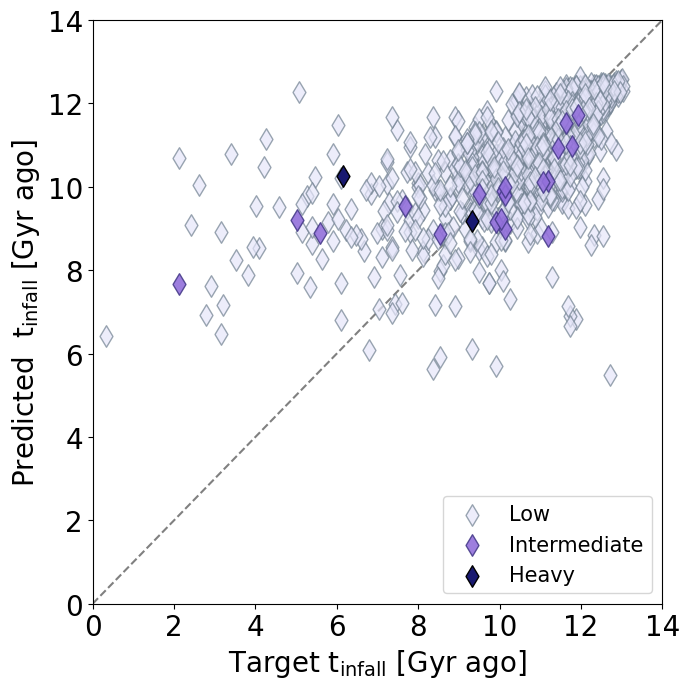}
 \end{center}
    \caption{Comparison between the target and predicted first infall times (in units of lookback Gyr) of satellite galaxies from \ASloth \nspace. This figure focuses on predicting the first infall time, unlike Figure \ref{fig:pred-targ}, which shows predictions for infall times to the MW-like halo galaxy. Each point represents a satellite galaxy, with colors indicating the mass groups defined in Table \ref{tab:data}. The model achieves an MSE loss of 1.66, demonstrating substantially higher accuracy compared to the MW-like halo infall time prediction model, which has an MSE of 5.04.
    \\
    Alt text: Predicted versus target values for the first infall times of \ASloth\ satellites, categorized by each mass group.}
    \label{fig:8}
\end{figure}

\section{Discussion}
\label{sec:4}

\subsection{The First Infall: the effect of group preprocessing}
\par 
In Section \ref{sss:mse}, we construct a model only using satellite galaxies that directly fell into a MW-like host galaxy, specifically excluding those that had belonged to any prior groups. Meanwhile, we find that $\tau_{90}$ shows a stronger correlation with the first infall time when satellites have experienced group preprocessing, as shown in Figure \ref{fig:f_i_M_i} and Table \ref{tab:mean_std}. Motivated by this, we now focus on galaxies that have undergone group preprocessing and train again our model using only those satellites to predict their actual first infall times prior to their accretion onto a MW-like host galaxy.

\par
Figure \ref{fig:8} shows the predicted first infall times for each mass group. We emphasize that Figure \ref{fig:8} focuses on the first infall of each satellite, unlike Figure~\ref{fig:pred-targ}, which displays the infall time to a MW-like host galaxy. The revised model yields an average MSE loss of 1.66, which is substantially lower than that of the default MW infall prediction model (5.04), excluding satellites that experience group preprocessing. This result indicates the significant influence of the first infall on the quenching of dwarf satellites (e.g., \citealp{Wetzel15, Simpson18}). It suggests that a large MSE with the default MW infall model may indicate prior group preprocessing. When we retrain the MW infall prediction model with a new set of 6,512 satellites identified as having undergone prior group preprocessing, the average MSE increases to 7.2. However, when the model is trained instead for the first infall prediction, the average MSE decreases to 1.66.

\par 
We also evaluate the model trained by satellites, experiencing group preprocessing, with the observational data. In our compiled observational samples, Car II, Hor I, and Ret II, are identified as members of the LMC group (\citealp{Battaglia22}). That is, these satellites were first accreted onto the LMC and then fell into the MW together. Since the first infall is expected to be the more crucial event for quenching than the MW infall event, we attempt to estimate their infall times of these satellites onto the LMC. Although no existing studies have provided LMC infall time estimates, our results can serve as a reference for future research. The MW infall times of Car II, Hor I, and Ret II are 7.69, 8.79, and 10.20 Gyr ago, respectively, which are in good agreement with F+19. The predicted infall times onto the LMC are 9.51, 9.46, and 10.30 Gyr ago, which are earlier than their infall times onto the MW, as expected.

\par 
Although these MW infall times are consistent with existing studies, the three satellites (Car II, Hor I, and Ret II) accreted onto the MW at different epochs (7.69, 8.79, and 10.20 Gyr ago). This appears to conflict with previous arguments that satellites from the same group should fall into the MW at a similar epoch (e.g., \citealp{Deason15, Wetzel15}). This discrepancy suggests a limitation of our MW infall prediction model, which is not optimized for satellites that undergo group preprocessing before falling into the MW. Accordingly, the model needs to be extended to incorporate the effects of group preprocessing explicitly in future work.

\par
However, it is important to note that the current conclusion is not statistically robust due to the limited number of known satellites with both reliable group membership and orbital histories. Therefore, discovering more satellites and reconstructing orbital histories based on their SFHs will be crucial for future comparisons. In this context, upcoming observational programs such as Gaia DR4 and DR5 (\citealp{Gaia451, Gaia452}) are expected to reveal additional dwarf satellites in the Local Group and provide their physical characteristics when combined with kinematic information from Gaia observations.

\subsection{Ionizing effect for the quenching of satellite galaxies}
\label{ss:ion}
\begin{figure}[tp]
 \begin{center}
  \includegraphics[width = 85mm]{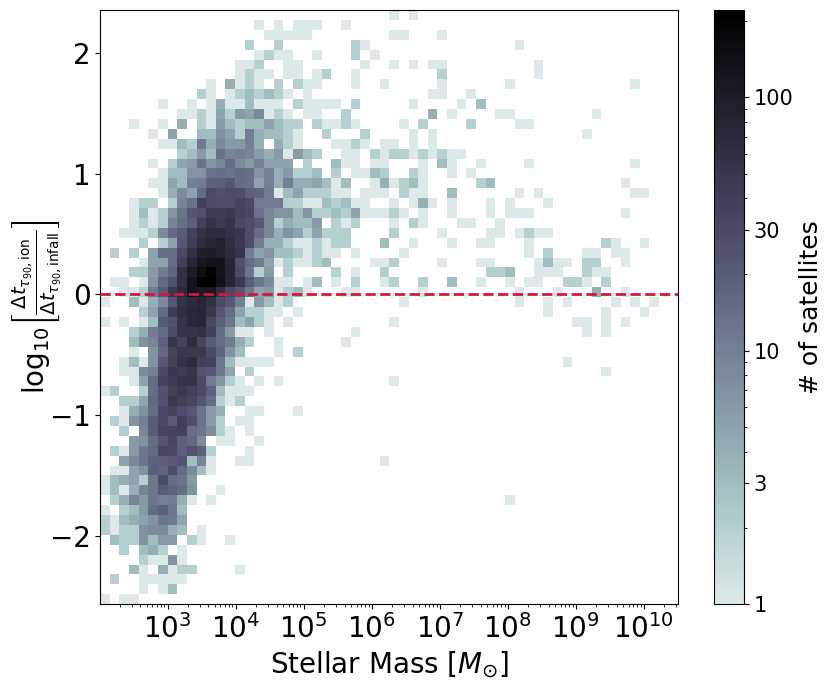}
 \end{center}
    \caption{A heatmap of the ratio between the quenching timescale of the first ionization and that of the first infall. Each cell indicates the number of satellites with a given stellar mass and quenching time ratio. A negative value implies that the ionizing effect is dominant, while a positive value indicates that infall dominates the quenching process. A sharp increase in satellites dominated by the ionizing effect is found below $10^4\ M_\odot$, which may correspond to low-mass satellites that quenched early but fell into their host galaxies at later times.
    
    Alt text: Predicted versus target values for the first infall times of \ASloth\ satellites, categorized by each mass group.}
    \label{fig:9}
\end{figure}
\begin{table}
\tbl{The quenching timescale (in Gyr) due to ionizing effects compared with the first infall event.}{
\begin{tabular}{c|c|c|c}
\hline
 & Low & Intermediate & Heavy \cr
\hline
Ionizing effect & 1.42 & 6.32 & 4.76 \cr
First infall & 4.29 & 3.46 & 4.67 \cr
\hline
\end{tabular}}
\label{tab:ion}
\end{table}

Cosmic reionization is another primary factor in the quenching of satellite galaxies (e.g., \citealp{Bullock2000, Brown14, Weisz14}). When early galaxies undergo active star formation, they collectively emit ionizing photons, creating expanding ionized bubbles that extend tens of kpc into the intergalactic medium (e.g., \citealp{Bovill2009, Tolstoy2009}). The elevated temperature ($\sim 10^4$ K) and pressure within these ionized regions inhibit gas cooling and accretion in low-mass halos, effectively preventing star formation in systems below the atomic cooling threshold. This global heating mechanism at high redshift acts on the halo scale rather than in localized star-forming regions, making it particularly effective at quenching dwarf galaxies in the early Universe.

\par
In \ASloth, the ionization status of a halo is determined as follows: a halo is considered fully ionized if it resides within an ionized bubble. However, if its halo mass exceeds the atomic cooling threshold ($T_{\rm vir}\gtrsim 10^4$ K), it is considered non-ionized regardless of its location, as the gas can cool efficiently as long as the ionizing background is not strong (e.g., \citealp{Oh2002, BrommLoeb2003, Visbal2017}). Nevertheless, if a halo is exposed to large fluxes of ionizing photons, $\zeta_{ion} = 6.7 \times 10^{6}$ photons $\mathrm{s^{-1}cm^{-2}}$, it is still classified as ionized. Lastly, when a halo mass of a satellite exceeds ten times the atomic cooling mass, the system remains non-ionized under any circumstances, being too massive to be affected by ionization. For a detailed description, see \citet{asloth}.

To determine the timing of ionization ($t_\mathrm{ion}$), we track each satellite in every simulation timestep and identify when it first enters an ionized state. We then compute the corresponding quenching timescale, $\Delta t_{\tau_{90},\ \mathrm{ion}}$, defined as the time difference between satellite ionization and quenching: $\Delta t_{\tau_{90},\ \mathrm{ion}} = t_\mathrm{ion}-\tau_{90}$. This allows us to compare which event—between infall and ionization—shows a stronger correlation with $\tau_{90}$. Until now, we have distinguished between the first infall (prior to MW infall) and MW infall in our models. However, to focus on comparisons with the first ionizing event, we now define the first infall as the initial infall experienced by a satellite, regardless of whether the host is the MW. 

Figure 9 shows a heatmap of ratio between $\Delta t_{\tau_{90},\ \mathrm{ion}}$ and $\Delta t_{\tau_{90},\ \mathrm{infall}}$. A negative value means that $\Delta t_{\tau_{90},\ \mathrm{infall}}$ is bigger, implying that the ionizing effect is dominant. On the other hand, a positive value indicates that the first infall is more dominant for the quenching of the satellite. Notably, for satellites with $M_{\star} \leq 10^4\ M_\odot$, ionization-dominated systems become increasingly common. Because such extremely low-mass satellites are highly sensitive to reionization, they are quenched very rapidly once exposed to ionizing radiation. This naturally contributes to the large scatter seen in the low-mass group in Figure \ref{fig:pred-targ} (left panel).

\par 

We further quantify the quenching timescales associated with the redefined first infall and the first ionizing event in Table~\ref{tab:ion}. The results are broadly consistent with the trends suggested in Figure \ref{fig:9}. In the heavy-mass group, 117 out of 156 satellites remain non-ionized at $z=0$.
 This occurs because, despite being located within ionized regions, heavy-mass satellites are likely to retain their neutral hydrogen gas due to their deep potential wells, which shield star-forming gas from external ionization. This indicates that heavy satellites are not significantly affected by physical processes such as an infall event or local reionization effects, resulting in a similar infall time prediction regardless of the ionization effect. For the intermediate-mass group, $\Delta t_{\tau_{90},\ \mathrm{ion}}$ is 6.32 Gyr, while $\Delta t_{\tau_{90},\ \mathrm{infall}}$ is 3.46 Gyr, indicating that quenching is more closely related to the infall event.

\par 

In contrast, the strongest link between $\tau_{90}$ and ionizing effects occurs in the lowest-mass regime. For satellites with $M_{\star} \lesssim 10^{4}\,M_{\odot}$, $\Delta t_{\tau_{90},\,\mathrm{ion}} = 1.42$ Gyr, which is about three times shorter than $\Delta t_{\tau_{90},\,\mathrm{infall}} = 4.29$ Gyr, confirming their extreme susceptibility to reionization, as suggested earlier. A transition is observed near $M_{\star} \sim 10^{4}\,M_{\odot}$, above which quenching becomes more strongly influenced by infall than by ionization. Using the 1,374 satellites with $M_{\star} = 10^{4}-10^{5}\,M_{\odot}$, we find $\Delta t_{\tau_{90,\,\mathrm{infall}}} = 3.75$ Gyr and $\Delta t_{\tau_{90,\,\mathrm{ion}}} = 4.3$ Gyr, indicating an infall-dominated quenching pathway. It is important to note the significant imbalance in sample sizes: satellites with $M_{\star} < 10^{4}\,M_{\odot}$ number 10,103, compared to only 1,374 with $M_{\star} > 10^{4}\,M_{\odot}$.

In summary, quenching in the heavy- and intermediate-mass groups is primarily driven by infall events, while ionizing effects become crucial for low-mass satellites. However, only the satellites with the lowest mass, $M_{\star} < 10^4\,\msun$, are dominated by ionization, whereas the more massive ones are more affected by infall.

\subsection{Caveat}
Generally, SAMs have several limitations compared to hydrodynamical simulations. \citet{Chen22} noted that mechanical and chemical feedback from Type Ia supernovae and UV background effect were not included in \ASloth. This omission may lead to extended star formation, resulting in smaller $\tau_{90}$ compared to models accounting for all necessary sub-grid physics. Many studies have emphasized that SAMs produce lower quenching fractions than hydrodynamic simulations. For example, \citet{Ayromlou21} compared the results of {\sc{L\_GALAXIES}} (LGal; \citealp{L-gal}), another SAM, with those of the hydrodynamical simulation of {\sc{IllustrisTNG}} (TNG; \citealp{TNG}). They demonstrated that galaxies are more likely to be quenched at high redshifts and have lower specific star formation rates (sSFR) in TNG compared to LGal. Similarly, \citet{Hirschmann12} compared a set of hydrodynamical simulations using the {\sc{gadget-2}} code with a SAM from \citet{s08}. They argued that SAMs often sustain excessive gas accretion and star formation due to underestimated feedback effects. In contrast, the feedback in hydrodynamical simulations is sufficiently strong to induce early quenching of galaxies, which matches more closely with observations.

\par 
The primary issue is that hydrodynamical simulations struggle to generate a large number of UFD samples. Currently, only a few high-resolution simulations, such as {\sc{DC Justice League}} simulation \citep{Applebaum21} and FIRE-2 simulation \citep{Hopkins18}, can resolve UFDs with $M_{\star} \leq 10^5\msun$ in the context of host galaxy environment. However, these state-of-the-art simulations require significant computational resources to solve complex equations and implement sophisticated sub-grid baryon physics. As a result, it remains challenging to generate a sufficiently large sample to train ML models.

\par
For example, \citet{Santistevan23} used the FIRE-2 simulation to study the orbital dynamics and histories of satellites with $M_{\star} \ge 10^4\msun$ around MW-mass hosts. They identified 473 satellites across 13 MW/M31-like hosts, which is insufficient for training compared to our SAM dataset of 12,170 satellites. To compensate for this limited sample size, they employed oversampling to reduce bias in their statistical analysis. However, oversampling can lead to overfitting, loss of statistical independence, or distortion of the underlying data distribution. Therefore, if future high-resolution hydrodynamic simulations, such as the upcoming simulations such as DARWIN (DAzzling Realization of dWarf galaxies in the next generation of cosmological hydrodynamic simulations) (Shin et al., in preparation), become capable of generating large samples of low-mass satellite galaxies, the performance and applicability of our ML model could be significantly improved.

\par 
Another consideration is the simplicity of our model. We use the default hyperparameters of {\sc Light}GBM, as they provide the best performance in our experiments. We find that further parameter tuning results in degraded performance. Also, we do not apply any feature extraction techniques, opting instead to use the physical quantities directly as input features. To ensure applicability to observational data, we exclude sequential quantities such as total star formation histories or mass evolutions, resulting in one-dimensional input features. Feature extraction techniques, such as convolutional neural networks or sequence encoders, are optimized for spatial or temporal patterns, which are not present in our tabular input. Applying such techniques is not only nontrivial but may also introduce unnecessary model complexity.

While our model's ability to predict MW infall times is valuable, it captures a genuine physical correlation between satellite galaxy quenching and infall event, rather than an arbitrary ML-derived relationship. Nonlinear transformations or latent representations can obscure direct physical insights, which are crucial in astrophysical analyses. Despite this advantage, the overall simplicity of our model, based on low-dimensional features and minimal tuning, can limit its predictive capacity. However, if future observational data provide time-dependent information, applying feature extraction to those sequential features could potentially enhance the model performance.

\par
Lastly, we focus exclusively on the first infall event, widely recognized as the dominant factor influencing quenching (e.g., \citealp{Wetzel15, Fillingham15}). Although many satellites have experienced multiple infalls, considering them would add unnecessary complexity to the analysis without clear benefits. Moreover, we have confirmed that the first infall is the most influential in quenching compared to subsequent infall events, as shown in Figure \ref{fig:f_i_M_i} and Table \ref{tab:mean_std}.


\section{Summary and Conclusion}

Understanding the infall time of satellite dwarf galaxies into larger systems is essential for uncovering the factors that determine their star formation histories. Despite its importance, identifying satellite infall times is not a straightforward task, often requiring the calculation of complete orbital histories through backward integration. However, based on the physical correlation between infall events and star formation quenching within dwarf galaxies, this study introduces a simplified method for estimating infall times. Specifically, we predict the infall times of satellite galaxies into Milky Way (MW)-like galaxies using the machine learning (ML) model \LightGBM, using three key features: quenching time ($\tau_{90}$), stellar mass ($M_{\star}$), and stellar metallicity ($[\mathrm{Fe}/\mathrm{H}]$). 

To train and construct the ML model, we obtain galaxy data from \ASloth, an advanced semi-analytic model that utilizes 30 dark matter-only MW-like galaxy simulations extracted from \Cater~project. During model construction, we account for preprocessing effects by classifying satellites into two categories: those that experienced prior group infall and those that directly accreted onto the MW-like host galaxy. Among a total of satellites, 6,512 underwent group preprocessing, while 5,590 directly fell into the MW-like host galaxy. Also, we divide satellites into three mass groups to assess which mass group exhibits a clearer correlation between infall time and quenching time.

The main findings of the tests with the data set randomly divided into training (80\%) and subsets of the test (20\%) are as follows.

\begin{itemize}

    \item Our default MW infall prediction model is constructed using only satellites that directly accreted onto the MW, as including group-preprocessed satellites degrades performance. By excluding these satellites, the model's performance improves, reducing the MSE from 6.92 to 5.04.

    \item For satellites in the low-mass group, the overall MSE is 5.07. Notably, this group exhibits a distinct trend based on infall time. Satellites that fell earlier than $\sim$ 4 Gyr ago show a significantly lower MSE of 2.54, while those that fell more recently exhibit a much higher MSE of 23.29. This contrast suggests that the recently infalling satellites in this low-mass group were likely quenched before accretion, weakening the correlation between infall time and quenching time.
    
    \item The intermediate-mass group yields the lowest MSE of 4.15, indicating a more consistent relationship between quenching and infall time across the satellites.
    
    \item The heavy-mass group exhibits the largest scatter, with an MSE of 5.28. This may reflect the reduced influence of environmental effects from the host on massive satellites, as their star formation continues regardless of infall events, thus weakening the link between their quenching epoch and infall time.
    
\end{itemize}

We then validate our model using the infall times of the MW satellites estimated from \citet{Fillingham19} (F+19) and \citet{Miyoshi20} (M+20), both of which provide infall time values based on the most widely used methods, such as backward integration and comparison with simulation results. We further compare our results with those of \citet{Barmentloo23} (B+23), where they also predicted satellite infall times using a ML approach. The main findings of this part are as follows.

\begin{itemize}

\item For 22 MW satellite galaxies, we compute MSE losses of 11.95 for F+19 and 18.63 for M+20, with some scatter attributed to differing methodologies for estimating infall times. Excluding outliers such as UMa I and CVn II significantly reduces the MSE for M+20, emphasizing the robust performance of our model for typical satellites, particularly in the low- and intermediate- mass groups.

\item Unlike F+19 and M+20, which do not use ML, B+23 is more directly comparable to our ML-based approach. Our model achieves a lower MSE than B+23 for F+19 in the low-mass group, but shows a higher MSE for M+20, likely due to discrepancies in CVn II values. For the intermediate-mass group, our model outperformed B+23 for both F+19 and M+20, while B+23 excels in the heavy-mass group.

\item We also predict the infall times of M31 satellites using our model predictions and compare these with the trends observed in {\sc A-Sloth} data. Our predictions show a significant correlation with $\tau_{90}$, with heavy-mass satellites generally falling into the M31 before quenching, while the low- and intermediate-mass groups show good agreement with $\tau_{90}$.

\end{itemize}

In addition to infall events, other physical mechanisms, particularly cosmic reionization, play a significant role in quenching star formation in dwarf galaxies, complicating the prediction of host infall times. We find that the quenching timescale due to ionization, denoted as $\Delta t_{\tau_{90},\ \mathrm{ion}}$, becomes significantly shorter for lower-mass satellites, with a sharp transition at $M_{\star} \sim 10^4~\msun$. Below this mass, ionization plays a crucial role in quenching compared to infall events.

Our study demonstrates that ML can effectively predict the infall times of satellite galaxies into MW-like host galaxies, particularly for low mass, those infalling to the MW before 4 Gyr ago, and for intermediate-mass populations. These galaxies frequently undergo quenching in the early Universe, a phase that traditional orbit integration methods find challenging to trace. Therefore, utilizing ML to estimate infall times could provide valuable insights. Future work incorporating more features and advanced ML techniques will be able to improve predictive accuracy. Moreover, access to more data from hydrodynamical simulations would allow us to train models that more accurately capture physical properties and this enhanced model will provide infall time information for observed satellite galaxies using upcoming telescopes.

\section*{Acknowledgements}

We express our gratitude to Hartwig Tilman for graciously sharing the \ASloth \nspace code with us, which is essential for this study. S. K. and M. J. is supported by grants from the National Research Foundation (NRF) funded by the Korean government (MSIT) (No.2021R1A2C109491713, No. 2022M3K3A1093827). S. K. is supported by the BK21 FOUR program through NRF under Ministry of Education (Kyung Hee University, School of Space Science).

\bibliography{myrefs}{}
\bibliographystyle{aasjournal}
\end{document}